\documentclass{llncs}
\let\llncssubparagraph\subparagraph
\let\subparagraph\paragraph
\usepackage[compact]{titlesec}
\let\subparagraph\llncssubparagraph

\usepackage{soul}
\usepackage{comment}
\usepackage[T1]{fontenc}
\makeatletter
\DeclareFontFamily{T1}{lcmtt}{\hyphenchar\font\m@ne}
\makeatother

\DeclareFontShape{T1}{lcmtt}{m}{n}{%
	<11.82><14.59><8><20.89><25.66><32.4><39.28>%
	ecltt8}{}

\usepackage{enumerate}
\usepackage{pdfpages}
\usepackage[latin1]{inputenc}
\usepackage{amsmath}
\usepackage{amsfonts}
\usepackage{amssymb}
\usepackage{graphicx}
\usepackage{epsfig}
\usepackage{epstopdf}
\usepackage{anyfontsize}
\usepackage{titlesec}
\titlespacing\section{1pt}{10pt plus 0pt minus 2pt}{1pt plus 2pt minus 2pt}
\titlespacing\subsection{1pt}{10pt plus 0pt minus 2pt}{0pt plus 2pt minus 2pt}
\titlespacing\subsubsection{1pt}{10pt plus 0pt minus 2pt}{pt plus 2pt minus 2pt}

\usepackage{subcaption}
\usepackage{enumitem}
\usepackage{multicol}
\usepackage{comment}
\usepackage{epsfig}
\usepackage{epstopdf}
\usepackage{color}
\usepackage{listings}
\usepackage{enumerate}
\usepackage{mathtools}
\usepackage{fancyvrb}
\usepackage[T1]{fontenc}
\makeatletter
\DeclareFontFamily{T1}{lcmtt}{\hyphenchar\font\m@ne}
\makeatother

\DeclareFontShape{T1}{lcmtt}{m}{n}{%
	<13.82><16.59><9><23.89><28.66><34.4><41.28>%
	ecltt8}{}

\usepackage{titlesec}
\usepackage{spverbatim}

\usepackage[normalem]{ulem}
\newenvironment{mycenter}[1][\topsep]
{\setlength{\topsep}{#1}\par\kern\topsep\centering}% \begin{mycenter}[<len>]
{\par\kern\topsep}% \end{mycenter}

\titlespacing*{\section}
{0pt}{1ex plus 0ex minus 0ex}{1ex plus 0ex minus 0.5ex}
\titlespacing*{\subsection}
{0pt}{1ex plus 0ex minus 0ex}{1ex plus 0ex minus 0.8ex}
\titlespacing*{\subsubsection}{0pt}{1ex plus 0pt minus 0.5ex}{1ex plus 0pt minus 0.5ex}

\definecolor{darkblue}{rgb}{0 0 0.9}
\definecolor{lightgray}{gray}{0.9}

\usepackage[bottom]{footmisc}
\usepackage{url}

\usepackage[colorlinks,allcolors=blue]{hyperref}
\title{Quantitative Corner Case Feature Analysis of Hybrid Automata with \textit{ForFET$^{SMT}$}}
\author{Antonio Anastasio Bruto da Costa\inst{1} \and Pallab Dasgupta\inst{1} \and Nikolaos Kekatos\inst{2}}
\institute{
	Dept. of Comp. Sci. and Engg., Indian Institute of Technology Kharagpur, India\\ 
	\and Verimag, Univ. Grenoble Alpes, France
}

\begin{document}
	
	\maketitle
	
	\begin{abstract}
		%This tool paper discusses the design and implementation of the formal {\em feature} evaluation tool for hybrid systems, {\em ForFET}. Features extend  the notion of assertions by associating a computable function to the match of an assertion. This paper illustrates the practical utility of feature evaluation through several examples.
		
		%This tool paper discuses the design and implementation of {\it ForFET$^{SMT}$} for evaluating quantitative property \textit{corners} for hybrid automata. ForFET$^{SMT}$ interacts with third-party tools, SpaceEx and SMT solver dReach/dReal. This paper proposes several standard quantitative property templates, illustrates some challenges of searching for feature corners using the SMT tool dReal and the utility of ForFET$^{SMT}$ through examples.
		
		The analysis and verification of hybrid automata (HA) models against rich formal properties can be a challenging task. Existing methods and tools can mainly reason whether a given property is satisfied or violated. However, such qualitative answers might not provide sufficient information about the model behaviors. This paper presents the \textit{ForFET$^{SMT}$} tool which can be used to reason quantitatively about such properties. It employs \emph{feature automata} and can evaluate quantitative property corners of HA. \textit{ForFET$^{SMT}$} uses two third-party formal verification tools as its backbone: the SpaceEx reachability tool and the SMT solver dReach/dReal. Herein, we describe the design and implementation of \textit{ForFET$^{SMT}$} and present its functionalities and modules. To improve the usability of the tool for non-expert users, we also provide a list of quantitative property templates.
		
		%We present \textit{ForFET$^{SMT}$,} a tool for the quantitative analysis of hybrid systems modeled as hybrid automata (HA). \textit{ForFET$^{SMT}$} makes use of \textit{features} to express quantitative measurements over behaviours of a HA and builds upon the tool ForFET. \textit{ForFET$^{SMT}$} is the first tool to provide a formal quantitative analysis of HA that provides the option to use both set-based reachability and SMT. In particular, it first computes \textit{feature} ranges formally over runs of a HA using the reachability tool SpaceEx.
		%Then, it refines the feature range using an expansion-bisection search using the delta-reachability SMT solver dReal. By leveraging SMT solvers, \textit{ForFET$^{SMT}$} acts as a practical tool, able to produce concrete traces associated with the extremal corners of the feature range. These traces can be used by experts for tuning the design to make the system more robust. 
	\end{abstract}

	%--------------------------------- INTRODUCTION -------------------------------------	
	\section{Introduction}
	Formal verification techniques can provide guarantees of correctness and performance for hybrid and cyber-physical systems. Nowadays, they are supported by several robust verification tools, e.g. {\tt SpaceEx~\cite{spaceex11}}, {\tt dReal}/{\tt dReach}~\cite{GaoKC13}. A common modeling formalism for the design of such systems is \emph{hybrid automata}~\cite{Alur01} (HA). HA  can exhibit non-deterministic behaviors and have been used to model control systems and analog mixed-signal circuit  designs~\cite{arch,dang04}.  
    Formalizing specifications of hybrid automata such that they can be verified automatically is not an easy task, especially in an industrial setting.
    
     There is a semantic mismatch between industrial requirements and formal specifications. Typically, industrial requirements are described in natural language\emph{"The caliper speed at contact must be below 2 mm/s"}, while formal specifications are expressed in a formal language, like  temporal logic (TL), e.g. $\square(q\rightarrow(\square p))$ with $p:=\{speed<=2mm/s\}$ and $q:=\{\text{caliper at contact}\}$. 
    
    Standard analysis tools~\cite{spaceex11} can answer reachability questions and can verify if given safety properties are satisfied. For more complex properties, one has to construct a monitor automaton and take its product with the HA~\cite{Nikolaos2018,kekatos2018formal}. In practice, however, the resulting automaton can be large, resulting in long analysis times and scalability issues. In addition, the answer provided by the tool is qualitative, i.e. yes or no. It is not possible to support quantitative measures, e.g. by what extent was the specification violated? In addition, describing common system properties requires the ability to express quantitative measures such as \textit{overshoot}, \textit{settling time}, or other \textit{timing} and \textit{value} quantities. 
    
    There are two directions to address this limitation. On the one hand, it is possible to use temporal logic. Much literature exists on TL, especially on Linear Temporal Logic (LTL)~\cite{Pnueli77}. Languages such as MITL~\cite{Alur96}, STL~\cite{porv01} and its extensions such as xSTL~\cite{XSTL_NickovicLMFU18} have been used for specifying specifications involving continuous signals. Some TL languages support the use of robustness metrics over properties~\cite{ROBUST_STL_DeshmukhDGJJS17}. Such metrics
	%Robustness metrics defined for an STL/MITL property 
	measure the \textit{distance} of runs of the system from unsafe regions defined by the property. %The parameterized version of STL (PSTL) allows temporal and predicate constants to be parameterized, transforming the quantitative analysis into parameter learning. 
	However, these languages are primarily designed to express specification correctness, and it can be tedious to use them to express quantitative measures.
    
    The other direction is to use \textit{features}~\cite{ain16}. Unlike temporal logics like MITL or STL, the language of features is designed to explicitly specify quantitative measures. The quantity is expressed as a computation resulting from matching a behaviour description.
    \textit{ForFET}~\cite{CostaD17} is a tool for computing an over-approximation for features, where the evaluation of a feature, written in the \textit{Feature Indented Assertion}~(FIA) language, over runs of a HA is automated.
    
     %Robustness metrics have also been defined over properties written in these languages~\cite{ROBUST_STL_DeshmukhDGJJS17}. Such metrics
	
	%Robustness metrics defined for an STL/MITL property measure the \textit{distance} of runs of the system from unsafe regions defined by the property. These metrics may be leveraged to express quantitative measures such as \textit{overshoot}, \textit{settling time}, or other \textit{timing} and \textit{value} quantities. %The parameterized version of STL (PSTL) allows temporal and predicate constants to be parameterized, transforming the quantitative analysis into parameter learning. 
	%However, these languages are primarily designed to express specification correctness, and it can be tedious to use them to express quantitative measures.
	
	%Unlike temporal logics like MITL, STL (or PSTL), the language of \textit{features}~\cite{ain16} is designed to explicitly specify quantitative measures. The quantity is expressed as a computation resulting from matching a behaviour description.

%\todo{Explain features}
%\todo{Explain feature ranges}
%\todo{Not a wrapper}

In a quantitative analysis, knowledge of the stimulus that produces the best and worst-case quantity (minimum or maximum) provides insight into the system and on how to modify the design to more robustly adhere to specifications.
	The tool \textit{ForFET$^{SMT}$} addresses  the needs described above, extending \textit{ForFET}, with the following:
	\begin{itemize}[noitemsep,topsep=0pt]
		\item Feature corner analysis using SMT. Using SMT has two advantages, it allows us to refine the feature range beyond what \textit{ForFET} produces, and also enables us to generate a witness trace describing the stimulus and behaviour for best and worst-case quantities.
		\item Support for parameterized features and an extended language for features having mixed urgent and non-urgent semantics.
		%\item Library of standard feature specification patterns.
		\item Usability and support: i) two translators, written in Matlab and Octave, for converting models from SpaceEx  formalism to \textit{ForFET}'s modeling language, ii) support for custom paths for workspace, models, and third-party tools.
	\end{itemize}
	
\section{Design and Implementation}

\begin{figure}[t]
\begin{multicols}{2}
\includegraphics[scale=0.75]{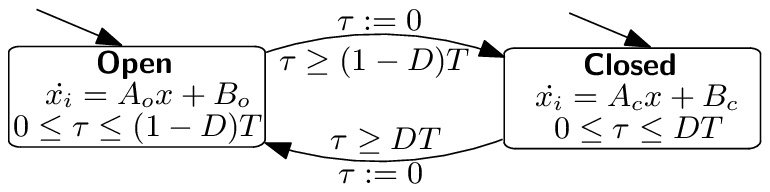}
\scriptsize{
\begin{Verbatim}[frame=single]
module buck(v,i,t)
    output v,i,t;	
    parameter 
        Vr = 12, 
        ... , 
        b1c = 0, 
        T = 1e-05,  
        D = 0.51667;	
    mode closed 
    begin
        ddt t = 1;
        ddt v = (a10c*i + a11c*v + b1c*Vs);
        ddt i = (a00c*i + a01c*v + b0c*Vs);
    end
    ...	
    property inv closed
        mode==closed |=> t<=D * T && t>=0;
    endproperty
\end{Verbatim}
}
\columnbreak
\scriptsize{
\begin{Verbatim}[frame=single]
    property trans closed_open
        mode==closed && mode'==open && 
        t>=D*T |=> i'==i && t'==0 && v'==v;
    endproperty
    ...	
    initial begin 
        set begin
            mode == closed;
            i == 0; v == 0; t == 0;
        end
    end
endmodule
\end{Verbatim}
}
\scriptsize{
\begin{Verbatim}[frame=single]
feature settlingTime(Vr,E);
begin
  var st;
  (v>=Vr+E) ##[0:$] 
  @+(state==Open) && (v<=Vr+E), st=$time 
  ##[0:$] @+(state==Open) && (v<=Vr+E)
    |-> settlingTime = st;
end
\end{Verbatim}
}
\end{multicols}
\caption{\small HA of the Buck Regulator, HASLAC Code-snippet of HA description, Quantitative Specification of Settling Time as a feature.}\label{fig:buck}
\end{figure}

We begin with a running example, as shown in Figure~\ref{fig:buck}, to explain the inputs to \textit{ForFET$^{SMT}.$} The example is of a buck regulator taken from standard benchmarks~\cite{buck}. The regulator receives an input voltage, and in the event of reasonably varying loads ensures that it provides an unchanging output voltage. 

\subsubsection{Hybrid Automaton Description in HASLAC}
An input to the tool is the HA description. The HA of the buck regulator has two locations, {\tt open} and {\tt closed}, indicating the state of the switch that charges the capacitor of the regulator.  

The description of the HA is specified in the \textit{Hybrid Automaton Specification Language for Analog Mixed-Signal (AMS) Circuits (HASLAC)}, as the model description language for \textit{ForFET} and \textit{ForFET$^{SMT}$}. \textit{HASLAC} is specially designed to mimic semiconductor circuit behavioural model description languages such as Verilong-AMS, to make adoption of formal analysis in the semiconductor circuit design flow less intimidating. \textit{HASLAC} describes each location of the HA as a {\tt mode}, with each transition and invariant expressed as a {\tt property}. In the model description, {\tt v} and {\tt i} are aliases for the HA variables ${\tt x_1}$ and ${\tt x_2}$. 

\subsubsection{Quantitative Specification using Features}

A \textit{feature} defines, formally, a quantitative specification, i.e. a measurement over behaviours of the system. Unlike properties, which either match or fail, having a Boolean outcome, the outcome of evaluating a feature is a real-valued interval. The language of features is easier to use and understand for non-experts, especially in the AMS domain, and it can be evaluated with the use of reachability tools.

In our example, the intent to \textit{measure} the time taken for the output voltage to settle into a stable state can be expressed as the feature {\tt settlingTime}. The feature contains three core components: (i) a set of behaviours over which measurements are made, (ii) variables, local to a feature, that may be assigned values in the antecedent, as a matching behaviour is observed, and (iii) the feature compute expression, over local variables, evaluated once the behaviour has matched. The behaviour described by the feature in Figure~\ref{fig:buck} reads as follows, \textit{"{\tt (v<=Vr+E)} is true and thereafter {\tt v} settles below {\tt (Vr+E)} for two successive openings of the capacitor switch"}. The expressions {\tt (v>=Vr+E)} and {\tt (v<=Vr+E)} are predicates over real-variables (PORVs). {\tt state} is a special variable allowing us to write predicates over the location labels of the HA. The construct {\tt @+(P)} represents an event, and is true only on the positive edge of the predicate {\tt P}. A behaviour in the feature expresses a sequence of Boolean expressions over PORVs and events separated by time-delays. The statement "{\tt P \#\#[a:b] Q}" is true whenever {\tt Q} occurs within a time interval of {\tt a} and {\tt b} from when {\tt P} is true; $a,b\in\mathbb{R}^+$, $b\geq a$. The syntax {\tt \#\#[a:b]} represents a time-delay. The symbol {\tt \$} represents the notion \textit{"anytime after a"}. Observe that {\tt P} can
be true over a dense time interval, and for each point in the interval where "{\tt P \#\#[a:b]}" is true, {\tt Q} can be true yielding an infinite number of matches. A more complete description of the language for features is available in~\cite{CostaTCAD18}.

\begin{remark}
A feature behaviour may match in one or multiple (potentially infinite) runs of the HA, at one or multiple (potentially infinite) time-points. Each match has the potential to yield a different feature value. Evaluating a feature over runs of a HA, therefore, yields an interval $[\mathcal{F}_{min},\mathcal{F}_{max}]$ of feature values. We call this a \textit{feature range}.
\end{remark}

\subsubsection{Algorithm}
\begin{figure}[t!]
\centering
	%\hspace{-5em}
	\begin{subfigure}[t]{0.6\textwidth}
		\centering
		\raisebox{-\height/2}{
		\includegraphics[width=\textwidth]{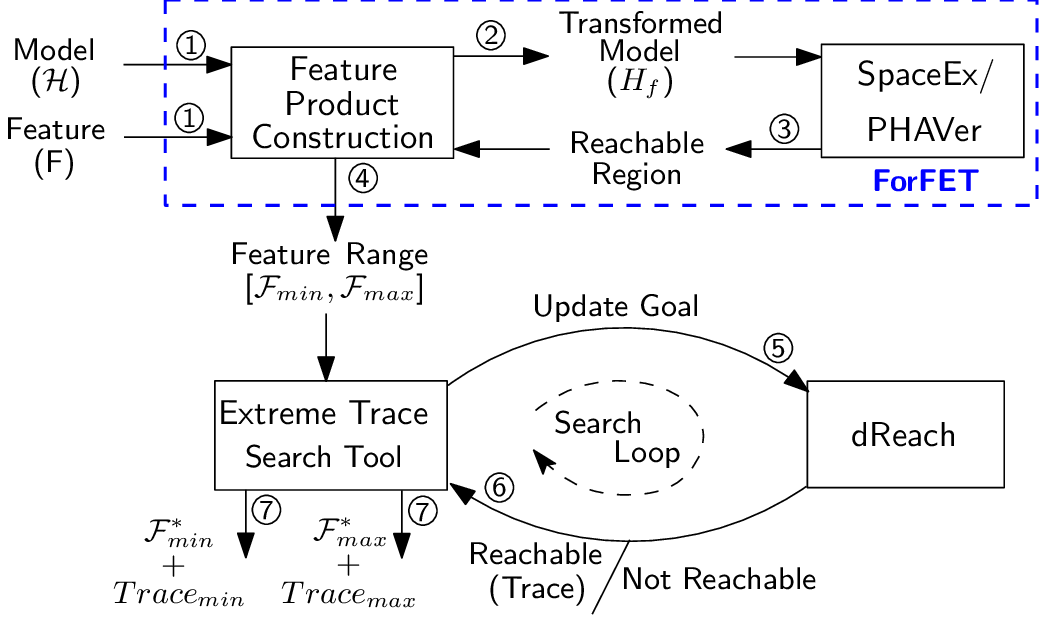}
		}
		\caption{Functional Overview}
		\label{fig:functionalOverview}
	\end{subfigure}
	~~
	\begin{subfigure}{0.35\textwidth}
	~\\~\\~\\
	\begin{subfigure}[t]{\textwidth}
		\centering
	    %\raisebox{-\height}{
	    \includegraphics[width=\textwidth]{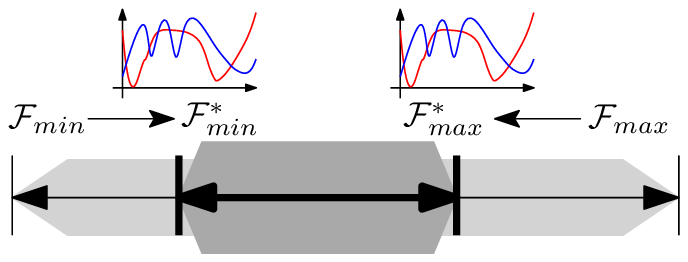}
	    %}
	   \caption{SMT Feature Refinement}
	   \label{fig:refinement}
	\end{subfigure}~\\~\\~\\~\\
	\begin{subfigure}[t]{\textwidth}
		\centering
        %\raisebox{-\height}{
        \includegraphics[width=\textwidth]{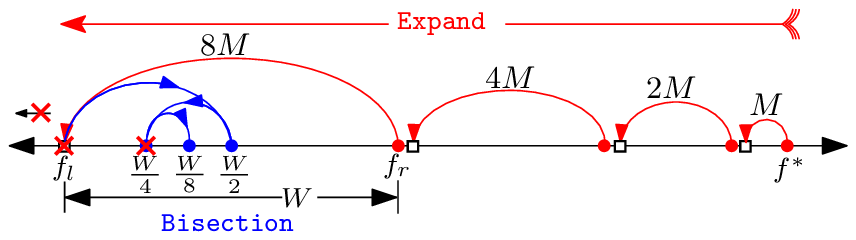}
        %}
        \caption{Computing the left corner}
        \label{fig:ETS-moves}
	\end{subfigure}
	\end{subfigure}
	\caption{\textit{ForFET$^{SMT}$}: {\em Corner Case Analysis}}\label{fig:ForFET-ETS}
\end{figure}

A functional overview of \textit{ForFET$^{SMT}$} is shown in Figure~\ref{fig:ForFET-ETS}. The tool \textit{ForFET} is marked within a blue box. \textit{ForFET$^{SMT}$} extends \textit{ForFET} by introducing an iterative refinement step that refines the range provided by \textit{ForFET}. It also introduces a wrapper around the SMT solver {\tt dReal} in order to correctly visualize a trace that acts as a witness for each corner of the feature range.

The tool works as follows. The user provides two inputs (\textit{Step 1}): a hybrid automaton model $\mathcal{H}$ and a feature specification  $\mathcal{F}$ (single or a set of features), \textit{ForFET$^{SMT}$} computes the product automaton (\textit{Step 2}) according to~\cite{CostaTCAD18}. \textit{Step~3} involves using {\tt SpaceEx}~\cite{spaceex11} to compute reach-sets for the transformed model $\mathcal{H}_F$. This results in a feature range $[\mathcal{F}_{min},\mathcal{F}_{max}]$ computed as an evaluation of the feature expression on the runs matching the feature sequence-expression (\textit{Step 4}). The feature range is refined iteratively through a search using an SMT solver (\textit{Steps 4 to 7}). 

{\tt HyST}~\cite{hyst_bak2015hscc} converter is used internally to translate the model $\mathcal{H}_f$ into an acceptable format for use with {\tt dReach}.  In each interaction, called a \textit{query}, between our tool and {\tt dReach}, a goal statement is constructed to direct {\tt dReach} to prove the existence/non-existence of a feature value in a given domain. Each query in \textit{Step 5} includes the model description for $\mathcal{H}_f$, a goal statement, and a maximum transition hop count $K$, which is translated by {\tt dReach} into SMT clauses. The response of the SMT solver (\textit{Step 6}) is either \textit{unsatisfiable} or \textit{satisfiable}. In the latter case, a single timed trace of the HA is made available. {\tt dReal} generates a trace as a JSON file with time-stamped valuations for the variables of the automaton, which is parsed to identify the feature values for the trace. The search concludes in \textit{Step 7} with a refined feature range $[\mathcal{F}^*_{min},\mathcal{F}^*_{max}]$ as well as a trace corresponding to each feature range corner value. 

	\subsubsection{Implementation}
	\textit{ForFET$^{SMT}$} is implemented in C/C++. The parsers for features and the HASLAC language are implemented in {\tt flex} and {\tt bison}, which are translated into C/C++. The language was chosen due to its efficiency for handling complex operations and data-structures involved in computing the product automaton of the HA and feature monitor.
	
	We represent the algorithm as a function which takes the HA and the feature as inputs, and produces a range of feature values $[\mathcal{F}^*_{min},\mathcal{F}^*_{max}]$ as output along with a trace that acts as a witness for each extremal corner of the range. The algorithm is guaranteed to terminate for bounded time traces~\cite{CostaTCAD18}. As such, it constitutes a procedure for computing feature ranges over bounded time horizons, which is expected in practice.
	
	\subsubsection{Challenges using {\tt dReal}}
	In general, HA use urgent locations to represent ordered discrete transformations. In a trace, {\tt dReal} provides a series of indexed time-ordered tuples representing a trace satisfying the query. In our experience, when the model $\mathcal{H}_f$ contains urgent locations, {\tt dReal} generates a {\tt NULL} tuple representing a visit to an urgent location. Visualization tools provided by the authors of {\tt dReal} do not support drawing traces containing a {\tt NULL} tuple. To enable visualization for all traces generated by {\tt dReal}, \textit{ForFET$^{SMT}$} post-processes traces generated by {\tt dReal}. It eliminates all {\tt NULL} tuples and re-indexes them to be consistent with the syntax expected by the visualization tool.
	
	\subsubsection{Support for SpaceEx Models}
	The SpaceEx modeling language has become the quasi-standard interchange format for defining and describing HA in the formal verification community~\cite{arch}. It offers a graphical user interface, respects the \textit{SX} grammar~\cite{cotton2010spaceex} and the models are written as XML files. \textit{ForFET$^{SMT}$} accepts HA models written in \textit{HASLAC}. To bridge this mismatch and facilitate the use of \textit{ForFET$^{SMT}$} with existing SpaceEx models and HA benchmarks, we provide two translators, written in MATLAB and Octave respectively. The translators require a SpaceEx model (necessary) and a configuration file (optional). They come with an XML parser (partly written in Java) and perform syntactic translation while also handling modeling differences. Note that there exist other converters tailored to hybrid automata and SpaceEx, e.g. {\tt HyST}~\cite{hyst_bak2015hscc}.
	
	\section{Installation and Usage}
	\subsubsection{Installation} 
	The \textit{ForFET$^{SMT}$} tool is available at a public GitHub repository. The repository may be cloned in full using the following command:
	\begin{mycenter}[5pt]
	\colorbox{lightgray}{\tt git clone \url{https://github.com/antoniobruto/ForFET2.git}}
	\end{mycenter}
	The tool is written in C/C++. Before building the tool, one needs to have {\tt g++}, {\tt flex}, {\tt bison}, {\tt glib-2.0}, {\tt json-glib-1.0} and C/C++ standard libraries for 32 bit binaries ({\tt ia32-libs} on Ubuntu 10.04 and later).
	
	The tool can be compiled by running {\tt ./buildForFET.sh} in the cloned directory. 
	One must also ensure that {\tt SpaceEx}\footnote{\url{http://spaceex.imag.fr/sites/default/files/downloads/private/spaceex\_exe-0.9.8f.tar.gz}} and the SMT translator and solver {\tt dReach}\footnote{\label{dRealLink}\url{https://github.com/dreal/dreal3/releases}} 
	and {\tt dReal}$^{\ref{dRealLink}}$ are installed and executable in the user's path. 
	The tool also uses the HA translator {\tt HyST} (provided with \textit{ForFET$^{SMT}$}) which requires a java run-time environment to be installed.
	
	\subsubsection{Usage}
	The compiled \textit{ForFET$^{SMT}$} tool can be interacted with through the command line. Once compiled, the tool binary resides within the {\tt forFET} directory as the binary {\tt forFET}. Standard invocation involves executing the binary with a configuration file, by running  {\tt ./forFET CONFIG-FILE-NAME}. The configuration file specifies where third party libraries may be found. An example configuration file is provided in {\tt forFET/default.cfg}. %ForFET$^{SMT}$ can be built both in macOS and Linux.; it have been tested macOS High Sierra and Catalina. However, limitations exist with the accompanied executable of SpaceEx and dReal/dReach.
	
	\section{Experimental Evaluation}
	
    In this Section, we present selected results on three models, provided with \textit{ForFET$^{SMT}$,} i.e. a battery charger, a cruise control, and a buck regulator. We tested a wide variety of features, capturing state-dependent, time-dependent, sequential-properties and combinations of them. Some of these properties can also be encoded as control specifications, e.g. overshoot or settling time. More details about the models, specifications, features, and analysis results can be found in the tool manual. For each model, the feature range is computed first using {\tt SpaceEx} and is then refined using SMT. Our observations show that, except in one case, reachability analysis and SMT require similar time to compute the expected feature range. The exception is when the model switches location often, like the Buck Regulator. In these cases, SMT might be more vulnerable to the state-space explosion. However, the additional computation overhead leads to tighter feature ranges.
	
	\section{Concluding Remarks}
	In this paper, we have presented the tool, \textit{ForFET$^{SMT}$}, that is a formal feature evaluation tool for hybrid automata (HA), emphasizing on its architecture and utilities. Features form a promising and practical research direction as they can be used on top of or alongside standard monitoring and hybrid reachability tools to provide quantitative measures about HA behaviors. \textit{ForFET$^{SMT}$} makes use of the HASLAC language for writing HA models and is linked to {\tt SpaceEx} reachability tool and {\tt dReal}/{\tt dReach} SMT solver. Using such an SMT solver to compute features can produce concrete traces for feature corner points and lead to the generation of tighter feature ranges. The traces for corners of the feature range can provide insights that guide experts to refine their designs.
	\clearpage
	\bibliographystyle{abbrv}
	%\nocite{*}
	%\bibliography{refs}
	\bibliography{references}	
\end{document}